\documentclass[aps,prd,reprint,groupedaddress,amsmath,amssymb,showpacs,
floatfix]{revtex4-1}
\usepackage{graphicx}
\usepackage{dcolumn}
\usepackage{bm}
\begin{document}
\title{An analysis of gravitational redshift from rotating body}
\author{Anuj Kumar Dubey}
\email[]{danuj67@gmail.com}
\affiliation{Department of Physics, Assam University, Silchar-788011, Assam, India.}
\author{A K Sen}
\email[]{asokesen@yahoo.com}
\affiliation{Department of Physics, Assam University, Silchar-788011, Assam, India.}
\date{\today}
\begin{abstract}
Gravitational redshift is  generally calculated without
considering the rotation of a body. Neglecting the rotation, the
geometry of space time can be described by using the spherically
symmetric Schwarzschild geometry. Rotation has great effect on
general relativity, which gives new challenges on gravitational
redshift. When rotation is taken into consideration spherical
symmetry is lost and off diagonal terms appear in the metric. The
geometry of space time can be then described by using the
solutions of Kerr family. In the present paper we discuss the
gravitational redshift for rotating body by using Kerr metric. The
numerical calculations has been done under Newtonian approximation
of angular momentum. It has been found that the value of
gravitational redshift is influenced by the direction of spin of
central body and also on the position (latitude) on the central
body at which the photon is emitted. The variation of
gravitational redshift from equatorial to non - equatorial region
has been calculated and its implications are discussed in detail.
\end{abstract}
\maketitle
\section{\label{sec:intro}Introduction}
 General relativity is not only relativistic theory of gravitation proposed by Einstein, but
 it is the simplest theory that is consistent with experimental data. Predictions of general
 relativity have been confirmed in all observations and experiments. Gravitational redshift of light
  is one of the predictions of general relativity and also provides evidence for the validity of the
  principle of equivalence. Any relativistic theory of gravitation consistent with the principle of
  equivalence will predict a redshift. \\ If, however, we observe on the Earth the spectrum emitted by
  the atoms located on the Sun, then, its lines appear to be shifted with respect to the lines of the
  same spectrum emitted on the Earth. Each line with frequency $\omega$ will be shifted through the
  interval $\Delta\omega$ given by formula:
  $\Delta\omega=\frac{\omega(V_{1}-V_{2})}{c^{2}}$
  where $V_{1}$ and $V_{2}$ are the potentials of the gravitational field at the points of emission and
  observation of the spectrum respectively (page 269, of Landau and Lifshitz [1]). If we observe on the earth
  a spectrum emitted on the sun or the stars, then magnitude of $V_{1}$ is greater than magnitude of $V_{2}$
   and it is clear that $\Delta\omega<$0, i.e, the shift occurs in the direction of lower frequency.
 The phenomenon we have described is called the \lq gravitational  redshift\rq.
\\ Adams in 1925 has claimed first about the confirmation of the predicted gravitational redshift from the measurement of the apparent radial velocity of Sirius B [2]. Pound and Rebka in 1959 were the first to experimentally verify the gravitational redshift from nuclear - resonance [3]. Pound and Snider in 1965 had performed an improved version of the experiment of Pound and Rebka, to measure the effect of gravity, making use of Mossbauer - Effect [4]. If the Source of radiation is at a height \textit{h} above an observer then the result found was (0.9990 $\pm$ 0.0076) times the value of $4.905\times
10^{-15}\times\frac{2gh}{c^{2}}$ predicted from principle of equivalence. Snider in 1972 has measured the redshift of the solar potassium absorption line at 7699 ${\AA}$ by using an atomic - beam resonance - scattering technique [5]. Krisher et al. in 1993 had measured the gravitational redshift of Sun [6].\\
Gravitational redshift has been reported by most of the authors without consideration of rotation of a body. Neglecting the rotation, the geometry of space time can be described using the well-known spherically symmetric Schwarzschild's geometry and information on the ratio $ \frac{M}{r}$ of a compact object can be obtained from the gravitational redshift, where M and r are mass and radius respectively. Thus the redshifted angular frequency ($\omega '$) and the original angular frequency ($\omega $) of a photon in Schwarzschild geometry are related by the relation (page 268, of Landau and Lifshitz [1])
\begin{equation}
\omega^{'}= \frac{\omega}{\sqrt{g_{tt}}}=\frac{\omega}{\sqrt{1-\frac{r_{g}}{r}}}
\end{equation}
where $r_g=(2GM/c^2)$ is the Schwarzschild radius.  P D Nunez and M Nowakowski in 2010 had obtained an expression for gravitational redshift factor of rotating body by using small perturbations to the Schwarzschild's geometry.
Two main results of this work include the derivation of a maximum angular velocity depending
only on the mass of the object and a possible estimates of the radius [7].
 \\\\ With this background the present paper is organized as follows. In Section -II, we derive
 the expression for four velocity and frame dragging in kerr field. In Section -III, we
 derive the expression for gravitational redshift factor ($\Re$) and redshift (Z)
 as the pulsar rotates. The derived expression of gravitational redshift is applied to
 calculate the values of gravitational redshift for various rotational bodies which includes
 the Sun and some millisecond pulsars. A pulsar as it rotates, the redshifted
 values of spectrum shows a periodicity. In Section -IV, we derive the expression for the coefficient of
 latitude dependence of redshift ($\kappa$). We further discuss its variation for different values of
 latitude at which photon is emitted. Finally, some conclusions are made in Section -V.
\section{\label{sec: Frame dragg} Frame dragging in kerr field}
\subsection{ Kerr Field}
When rotation is taken into consideration spherical symmetry is
lost and off - diagonal terms appear in the metric and the most
useful form of the solution of Kerr family is given in terms of t,
r, $\theta$ and $\phi$, where t, and r are Boyer - Lindquist
coordinates running from - $\infty$ to + $\infty$, $\theta$  and
$\phi$, are ordinary spherical coordinates in which  $\phi$ is
periodic with period of 2 $\pi $ and $\theta$ runs from 0 to
$\pi$.
\\Covariant form of metric tensor with signature (+,-,-,-) is expressed as
\begin{equation}
ds^{2} = g_{tt}c^{2}dt^{2}+ g_{rr} dr^{2} + g_{\theta\theta} d\theta^{2} +g_{\phi\phi} d\phi^{2} + 2 g_{t \phi} c dt d\phi
\end{equation}
\\ Non-zero components $g_{ij} $ of Kerr family are given as follows (page 346, of Landau and Lifshitz [1] and page 261-263, of Carroll [8]),
\begin{equation}
 g_{tt}=\frac{\Delta-a^{2}sin^{2}\theta}{\rho^{2}} = (1-\frac{r_{g} r}{\rho^{2}})
\end{equation}
\begin{equation}
g_{rr}=-\frac{\rho^{2}}{\Delta}
\end{equation}
 \begin{equation}g_{\theta\theta}=-{\rho^{2}}
 \end{equation}
\begin{widetext}
\begin{equation}
g_{\phi\phi}= -\frac{[(r^{2}+a^{2})^{2}-a^{2}\Delta sin^{2}\theta]sin^{2}\theta}{\rho^{2}}=-[r^{2}+a^{2}+\dfrac{r_{g} r a^{2}sin^{2}\theta}{\rho^{2}}]sin^{2}\theta
 \end{equation}
 \end{widetext}
\begin{equation}
g_{t\phi}=\frac{a sin^{2}\theta (r_{g}r-e^{2})}{\rho^{2}}
\end{equation}
with
\begin{equation}
\rho^{2} = r^{2}+ a^{2}cos^{2}\theta
\end{equation}
\begin{equation}
 \Delta =r^{2}-r_{g}r + a^{2}+e^{2}
\end{equation}
where the parameters  e and $ a (= \frac{J}{Mc})$ are respectively charge and rotation parameter of the source.
J is the angular momentum of the compact object or central body, which can be also written as $J = I\Omega$. M,
I and $\Omega$ are the mass, moment of inertia and angular velocity of the central body respectively. Charge (e) is written as $ e = \sqrt {Q^{2}+P^{2}} $ , where Q is electric charge and P is magnetic charge. In case of Kerr solution in 1963 both Q and P vanishes [9], for Kerr - Newman solution in 1965 only P vanish [10] but in case of Kerr - Newman - Kasuya solution in 1982 both Q and P are non-vanishing [11].\\
The values of $g_{tt}$, $g_{t\phi}$ and $g_{\phi\phi}$ for polar and equatorial regions can be obtained from equations (3) to (9).
 \\ For equatorial plane where $\theta =\frac{\pi}{2} $, from equations (8) and (9), if we consider charge e = 0, then $ \rho^{2} = r^{2}$  and $  \Delta = r^{2}-r_{g}r + a^{2} $.
 \\ From equations (3), (6) and (7) metric elements $ g_{tt}$, $ g_{\phi\phi}$ and $g_{t\phi}$ for Kerr space time at equator becomes,
 \begin{equation}
 g_{tt}(\theta =\frac{\pi}{2})= (1-\frac{r_{g}}{r})
\end{equation}
\begin{equation}
 g_{\phi\phi}(\theta =\frac{\pi}{2})= -(r^{2}+a^{2}+\frac{r_{g}a^{2}}{r})
\end{equation}
\begin{equation}
g_{t\phi}(\theta =\frac{\pi}{2}) =(\frac{r_{g}a}{r})
\end{equation}
\\ At poles where $\theta =0 $, from equations (8) and (9), if we consider charge e = 0,
 then $ \rho^{2} = r^{2}+a^{2}$  and $ \Delta =r^{2}-r_{g}r + a^{2} $.
\\ From equations (3), (6) and (7) metric elements $ g_{tt}$, $ g_{\phi\phi}$ and $g_{t\phi}$ for Kerr space
time at poles become,
\begin{equation}
 g_{tt}(\theta=0)= 1-\frac{r_{g}r}{r^{2}+a^{2}}
\end{equation}
\begin{equation}
g_{\phi\phi}(\theta=0)= 0
\end{equation}
\begin{equation}
g_{t\phi}(\theta=0)= 0
\end{equation}
\subsection{Four - Velocity in Kerr Field}
Four - dimensional velocity (four - velocity) of a particle is defined as the four - vector (page 23, of Landau
and Lifshitz [1])
\begin{equation}
u^{i} =\frac{dx^{i}}{ds}
\end{equation}
 where the index i takes on the values 0,1,2,3 and $x^{0} = ct$, $x^{1} = r$, $x^{2} = \theta$, $x^{3} = \phi$.
 \\ Using equation (2) we can write
 \begin{widetext}
 \begin{equation}
ds =\sqrt{g_{tt}c^{2}dt^{2}+ g_{rr} dr^{2} + g_{\theta\theta} d\theta^{2} +g_{\phi\phi} d\phi^{2} + 2 g_{t \phi} c dt d\phi}
\end{equation}
The components of four - velocity are
\begin{equation}
 u^{0}\equiv u^{t} =\frac{dx^{o}}{ds} = \frac{cdt}{\sqrt{g_{tt}c^{2}dt^{2}+ g_{rr} dr^{2} + g_{\theta\theta} d\theta^{2} +g_{\phi\phi} d\phi^{2} + 2 g_{t \phi} c dt d\phi}}
\end{equation}
\begin{equation}
 u^{1}\equiv u^{r} =\frac{dx^{1}}{ds}  = \frac{dr}{\sqrt{g_{tt}c^{2}dt^{2}+ g_{rr} dr^{2} + g_{\theta\theta} d\theta^{2} +g_{\phi\phi} d\phi^{2} + 2 g_{t \phi} c dt d\phi}}
\end{equation}
\begin{equation}
 u^{2}\equiv u^{\theta} =\frac{dx^{2}}{ds} = \frac{d\theta}{\sqrt{g_{tt}c^{2}dt^{2}+ g_{rr} dr^{2} + g_{\theta\theta} d\theta^{2} +g_{\phi\phi} d\phi^{2} + 2 g_{t \phi} c dt d\phi}}
\end{equation}
\begin{equation}
 u^{3}\equiv u^{\phi} =\frac{dx^{3}}{ds} = \frac{d\phi}{\sqrt{g_{tt}c^{2}dt^{2}+ g_{rr} dr^{2} + g_{\theta\theta} d\theta^{2} +g_{\phi\phi} d\phi^{2} + 2 g_{t \phi} c dt d\phi}}
\end{equation}
From equations (18) and (21) we can write
\begin{equation}
\frac{u^{3}}{u^{0}}\equiv\frac{u^{\phi}}{u^{t}} = \frac{d\phi}{c dt}
\end{equation}
\end{widetext}
For a sphere, the photon is emitted at a location on its surface where $dr = d\theta =0$, when the sphere rotates.
\\So the expressions for four - velocity becomes,
 \begin{equation}
 u^{0}\equiv u^{t} =\frac{1}{\sqrt{g_{tt}+ g_{\phi\phi}(\frac{d\phi}{c dt})^{2} +2 g_{t \phi}(\frac{d\phi}{c dt})}}
\end{equation}
\begin{equation}
u^{1}\equiv u^{r}=0
\end{equation}
\begin{equation}
u^{2}\equiv u^{\theta} = 0
\end{equation}
\begin{equation}
 u^{3}\equiv u^{\phi} = u^{t} \frac{d\phi}{c dt}
\end{equation}
Hence four - velocity of an object in Kerr field can be expressed as,
\begin{equation}
u^{i} = (u^{t},0,0,u^{t} \frac{d\phi}{c dt})
\end{equation}
\subsection{Frame Dragging in Kerr Field}
The Lagrangian $\pounds$ of the test particle (\textbf{with mass $m$}) can be expressed as
\begin{equation}
\pounds  = \frac{g_{ij}\dot{x}^{i}\dot{x}^{j}}{2}
\end{equation}
(Dot over a symbol denotes ordinary differentiation with respect to an affine parameter $\xi$)\\
From the Lagrangian of the test particle, we can obtain the momentum of the test particle as,
\begin{equation}
P_{i} = \frac{\partial\pounds}{\partial\dot{x}^{i}}= g_{ij} \dot{x}^{j}
\end{equation}
Thus corresponding momentum in the coordinates of t, r, $\theta$, and $\phi$ are given by
\begin{equation}
P_{t} \equiv - E= c^{2} g_{tt} \dot{t} + c g_{t\phi} \dot{\phi}
\end{equation}
\begin{equation}
P_{r} = g_{rr} \dot{r}
\end{equation}
\begin{equation}
P_{\theta} = g_{\theta\theta} \dot{\theta}
\end{equation}
\begin{equation}
P_{\phi} \equiv L= c g_{t\phi} \dot{t} + g_{\phi\phi} \dot{\phi}
\end{equation}
Since the space time of the Kerr family is stationary and axially symmetric, the momenta $P_{t}$ and $P_{\phi}$  are conserved along the geodesics. So we obtain two constants of motion: one is corresponding to the conservation of energy (E) and the other is the angular momentum (L) about the symmetry axis.\\
we can write from equations (30) and (33)
\begin{equation}
c\dot{t} =\frac{g_{\phi\phi} E + c g_{t\phi} L }{c g^{2}_{t\phi}- c g_{tt} g_{\phi\phi}}
\end{equation}
and
\begin{equation}
\dot{\phi} = -\frac{g_{t\phi} E + c g_{tt} L }{c g^{2}_{t\phi}- c g_{tt} g_{\phi\phi}}
\end{equation}
Using equations (34) and (35), we can obtain  $\frac{d\phi}{cdt}$
\begin{equation}
\frac{\dot{\phi}}{c\dot{t}}=\frac{u^{\phi}}{u^{t}}= \frac{d\phi}{cdt} =- \frac{g_{t\phi} E + c g_{tt} L }{g_{\phi\phi} E + c g_{t\phi} L}
\end{equation}
Substituting the value of $g_{tt}$, $g_{\phi\phi}$, and $g_{t\phi}$ from equations (3), (6) and (7) in above equation (36),
$$\frac{d\phi}{cdt}= -\frac{\frac{a sin^{2}\theta r_{g}r}{\rho^{2}} E + c(1-\frac{r_{g} r}{\rho^{2}}) L }{-(r^{2}+a^{2}+\dfrac{r_{g} r a^{2}sin^{2}\theta}{\rho^{2}})sin^{2}\theta E + c \frac{a sin^{2}\theta r_{g}r}{\rho^{2}} L}$$
\begin{equation}
 \frac{d\phi}{cdt}=\frac{\frac{c L}{sin^{2}\theta} (1-\frac{r_{g}r}{\rho^{2}})+\frac{r_{g}ra}{\rho^{2}} E}{-\frac{ c r_{g}r a}{\rho^{2}} L + E (r^{2}+a^{2}+\frac{r_{g}r a^{2}}{ \rho^{2}}sin^{2}\theta)}
\end{equation}
Alternatively following Landau and Lifshitz ([1] page 351), we can also find $\frac{d\phi}{cdt}$ in the following way:\\
The four momentum of the test particle (with mass $m$) is
\begin{equation}
p^{i}= m \frac{dx^{i}}{ds}
\end{equation}
\begin{equation}
m\frac{cdt}{ds}= -\frac{r_{g}r a}{\rho^{2}\Delta} c L +\frac{E}{\Delta} (r^{2}+a^{2}+\frac{r_{g}r a^{2}}{ \rho^{2}}sin^{2}\theta)
\end{equation}
\begin{equation}
m\frac{d\phi}{ds}= \frac{c L}{\Delta sin^{2}\theta} (1-\frac{r_{g}r}{\rho^{2}})+\frac{r_{g}ra}{\rho^{2}\Delta} E
\end{equation}
From above equations (39) and (40), we can find $\frac{d\phi}{cdt}$ as
$$\frac{d\phi}{cdt} = \frac{\frac{c L}{sin^{2}\theta} (1-\frac{r_{g}r}{\rho^{2}})+\frac{r_{g}ra}{\rho^{2}} E}{-\frac{r_{g}r a}{\rho^{2}} c L + E (r^{2}+a^{2}+\frac{r_{g}r a^{2}}{ \rho^{2}}sin^{2}\theta)}$$
The above expression is identical to the expression (37) obtained earlier, where E and L are conserved energy and components of angular momentum along the axis of symmetry of the field respectively.\\
Frame dragging is a general relativistic feature of all solutions
to the Einstein field equations associated with rotating masses.
Due to the influence of gravity frame dragging or dragging of
inertial frame arises in the Kerr metric. $\frac{d\phi}{cdt}$ is
termed as angular velocity of frame dragging [12]. The
equation (37) is the general expression of angular
velocity of frame dragging ($\frac{d\phi}{cdt}$) in Kerr field.
\\ Following (Cunningham and Bardeen 1972 [13]), the photon trajectory is independent of its energy and may be
described by the parameter $\frac{L}{E}$, which is related to the direction cosines of a beam of radiation with
respect to the $\phi$ - direction and $\theta$ - direction in the local rest frame.
\\In case of photon the energy (E) and linear momentum (p) are expressed as
\begin{equation}
E^{2} = p^{2}c^{2}+m_{0}^{2}c^{4}
\end{equation}
where $m_{0}$ is rest mass of the photon which is zero. So above equation becomes
\begin{equation}
E = pc
\end{equation}
\\The angular momentum (L) about the symmetry axis may be expressed as
\begin{equation}
L = p b
\end{equation}
We may note that the  symbol \lq $\pounds$\rq \ stands for Lagrangian and \lq L\rq \ for angular momentum about the symmetry axis. \\
It is assumed that b$\ggg r_{g}$ and b $\ggg a$. Here b is perpendicular distance between the axis of rotation of the central body and direction of ray of light.\\
Considering the ray of light which is emitted radially outward
from the surface of the star (Pulsar or Sun) with radius \lq R\rq,
we can substitute, $b= R sin\phi sin\theta$ in above equation (43)
and finally we can write
\begin{equation}
L = p R sin\phi sin\theta
\end{equation}
where angle $\phi=\Omega t$, is measured from the azimuthal axis , which is defined as the direction when
source of light (on the spherical surface) is nearest to the observer.\\
$\Omega= \frac{2\pi}{T}$, where T is period of rotation of star (Pulsar or Sun).\\
Substituting the value of E and L from equations (42) and (44) in the expression (37) of $\frac{d\phi}{cdt}$,
\begin{widetext}
\begin{equation}
\frac{d\phi}{cdt} = \frac{\frac{pc R sin\phi sin\theta}{sin^{2}\theta} (1-\frac{r_{g}r}{\rho^{2}})+\frac{r_{g}ra}{\rho^{2}} (pc)}{-\frac{r_{g}r a}{\rho^{2}} (pc R sin\phi sin\theta) + (pc) (r^{2}+a^{2}+\frac{r_{g}r a^{2}}{ \rho^{2}}sin^{2}\theta)}
\end{equation}
\begin{equation}
\frac{d\phi}{cdt}( \phi, \theta)=\frac{\frac{R sin\phi sin\theta}{sin^{2}\theta} (1-\frac{r_{g}r}{\rho^{2}})+\frac{r_{g}ra}{\rho^{2}}}{-\frac{r_{g}r a}{\rho^{2}} (R sin \phi sin\theta) + (r^{2}+a^{2}+\frac{r_{g}r a^{2}}{ \rho^{2}}sin^{2}\theta)}
\end{equation}
\end{widetext}
 For equatorial plane where $\theta =\frac{\pi}{2} $ and $ \rho^{2} = r^{2}$, the above expression (46) of $\frac{d\phi}{cdt}$ can be written as:
\begin{equation}
\frac{d\phi}{cdt}(\phi,\theta =\frac{\pi}{2}) = \frac { (1-\frac{r_{g}}{r})( R sin\phi)+\frac{r_{g}a}{r} }{-\frac{r_{g} a}{r} ( R sin\phi) + (r^{2}+a^{2}+\frac{r_{g} a^{2}}{r})}
\end{equation}
The above expression (47) is the expression of frame dragging
$(\frac{d\phi}{cdt})$ on the  equatorial plane.
\\ Again for any general $\theta$ and at $\phi=\frac{\pi}{2}$, the expression for $\frac{d\phi}{cdt}$ from equation (46) can be written as:
\begin{equation}
\frac{d\phi}{cdt}(\phi =\frac{\pi}{2},\theta)=\frac{\frac{R}{sin\theta} (1-\frac{r_{g}r}{\rho^{2}})+\frac{r_{g}ra}{\rho^{2}}}{-\frac{r_{g}r a}{\rho^{2}} (R sin\theta) + (r^{2}+a^{2}+\frac{r_{g}r a^{2}}{ \rho^{2}}sin^{2}\theta)}
\end{equation}
 Substituting $\theta =\frac{\pi}{2} $ and $ \rho^{2} = r^{2}$, the above expression (48) of $\frac{d\phi}{cdt}$ can be written as:
\begin{equation}
\frac{d\phi}{cdt}( \phi =\frac{\pi}{2},\theta =\frac{\pi}{2})=\frac{R (1-\frac{r_{g}}{r})+\frac{r_{g}a}{r}}{-\frac{r_{g} a}{r} R + (r^{2}+a^{2}+\frac{r_{g} a^{2}}{ r})}
\end{equation}
\section{\label{sec:Grav redshift} Gravitational redshift from rotating body}
\subsection{Frequency in Kerr Field}
Let f be any quantity describing the field of the wave. For a plane monochromatic wave f has the form (page 140,
 of Landau and Lifshitz [1])
\begin{equation}
f= a e^{\textbf{j}(k.r-\omega t +\alpha)} =  a e^{\textbf{j}(k_{i}x^{i} +\alpha)}
\end{equation}
where $\textbf{j} = \sqrt{-1}$, and i used as superscript and
subscript with indices i=0,1,2,3.
\\k is propagation constant and $k_{i}$ is the wave four - vector.
\\We write the expression for the field in the form
\begin{equation}
f = a e^{\textbf{j}\Psi}
 \end{equation}
where $\Psi\equiv -k_{i}x^{i} +\alpha$, is defined as eikonal.
\\Over small region of space and time intervals the eikonal $\Psi$ can be expanded in series
 to terms of first order, we have
\begin{equation}
\Psi=\Psi_{0}+r.\frac{\partial\Psi}{\partial r}+t\frac{\partial\Psi}{\partial t}
\end{equation}
As a result one can write (page 141, of Landau and Lifshitz [1])
\begin{equation}
k_{i}=-\frac{\partial\Psi}{\partial x^{i}}
\end{equation}
where $k_{i}$ is the wave four - vector and the components of the four - wave vector $k_{i}$ are related by
\begin{equation}
{k_{i}}{k^{i}} = 0
\end{equation}
or
\begin{equation}
\frac{\partial\Psi}{\partial x^{i}}\frac{\partial\Psi}{\partial x_{i}}=0
\end{equation}
This equation called the eikonal equation is fundamental equation of geometrical optics.\\
 Using equation (53), the components of wave four - vector are
\begin{equation}
k_{0}=-\frac{\partial\Psi}{\partial x^{0}}= -\frac{\partial\Psi}{\partial ct}=\frac{\omega}{c}
\end{equation}
\begin{equation}
k_{1}=k_{x}= -\frac{\partial\Psi}{\partial x^{1}}= -\frac{\partial\Psi}{\partial x}
\end{equation}
\begin{equation}
k_{2}=k_{y}= -\frac{\partial\Psi}{\partial x^{2}}= -\frac{\partial\Psi}{\partial y}
\end{equation}
\begin{equation}
k_{3}=k_{z}= -\frac{\partial\Psi}{\partial x^{3}}= -\frac{\partial\Psi}{\partial z}
\end{equation}
Hence we can write the wave - four vector
 \begin{equation}
k_{i}= (k_{0},k_{1},k_{2},k_{3})=( \frac{\omega}{c},k_{x},k_{y},k_{z})
\end{equation}
$k_{0}(=\frac{\omega}{c})$ is the time component of four - wave vector and this frequency is measured in
terms of the world time.
\\Frequency measured in terms of proper time ($\tau$) is defined as
\begin{equation}
\omega^{'}=-\frac{\partial\Psi}{\partial \tau}
\end{equation}
This frequency $ \omega^{'} $ is different at different point of space.
\begin{equation}
\omega^{'}= -\frac{\partial\Psi}{\partial x^{0}}\frac{\partial x^{0}}{\partial \tau} = \omega \frac{\partial x^{0}}{c\partial \tau} =\omega u^{0}
\end{equation}
as $\frac{\partial x^{0}}{c\partial \tau}$ is nothing but $u^{0}$.\\
Substituting the value of $ u^{0} $ in above equation (62) from equation (23)
\begin{equation}
\omega^{'}=\frac{\omega}{\sqrt{g_{tt}+ g_{\phi\phi}(\frac{d\phi}{c dt})^{2} +2 g_{t \phi}(\frac{d\phi}{c dt})}}
\end{equation}
This $\omega^{'}$ is the frequency measured by a distant observer in terms of proper time ($\tau$) and $\omega$ is the frequency measured in terms of the world time (t).\\
\\\textbf{Alternative method: Frequency in Kerr field}
\\\\ Any observer measures the frequency ($\omega^{'}$) of a photon
following the null geodesic $x^{i} (\xi)$ which can be calculated
by the expression given as (page 217 of Carroll 2004 [8] and page
108 of Straumann 1984 [14],
\begin{equation}
\omega^{'} = u^{i}\frac{dx_{i}}{d\xi}=u^{i} g_{ij}\frac{dx^{j}}{d\xi}
\end{equation}
\begin{equation}
\omega^{'} = u^{t}(g_{tt} c^{2}\dot{t} + c g_{t\phi} \dot{\phi})+u^{\phi}(g_{t\phi} c \dot{t} + g_{\phi\phi} \dot{\phi})
\end{equation}
Using equation (30) and (33) in the above equation (65), we can
write
\begin{equation}
\omega^{'} = u^{t}(-E) + u^{\phi}(L)
\end{equation}
Again using equation (26) in above equation (66) we can write:
\begin{equation}
\omega^{'} = u^{t}(-E + \frac{d\phi}{c dt} L)
\end{equation}
or
$$\omega^{'} = \omega u^{0}$$
The above expression is identical to the expression (62) obtained earlier.
Here we define $(-E + \frac{d\phi}{c dt} L)$ as $\omega$, where $\omega$ can be identified as the frequency
corresponding to world time as in equation (56) and equation (62).
\\ As a result, we can write the expression of frequency measured in terms of proper time or as observed by distant observer as:
$$\omega^{'}=\frac{\omega}{\sqrt{g_{tt}+ g_{\phi\phi}(\frac{d\phi}{c dt})^{2} +2 g_{t \phi}(\frac{d\phi}{c dt})}}$$
The above expression is identical to the expression (63) obtained earlier. Thus the two approaches give us identical expression for frequency observed by distant observer.
\subsection{Gravitational redshift factor and redshift}
\begin{figure*}
\includegraphics[width=50pc, height=40pc,angle=270]{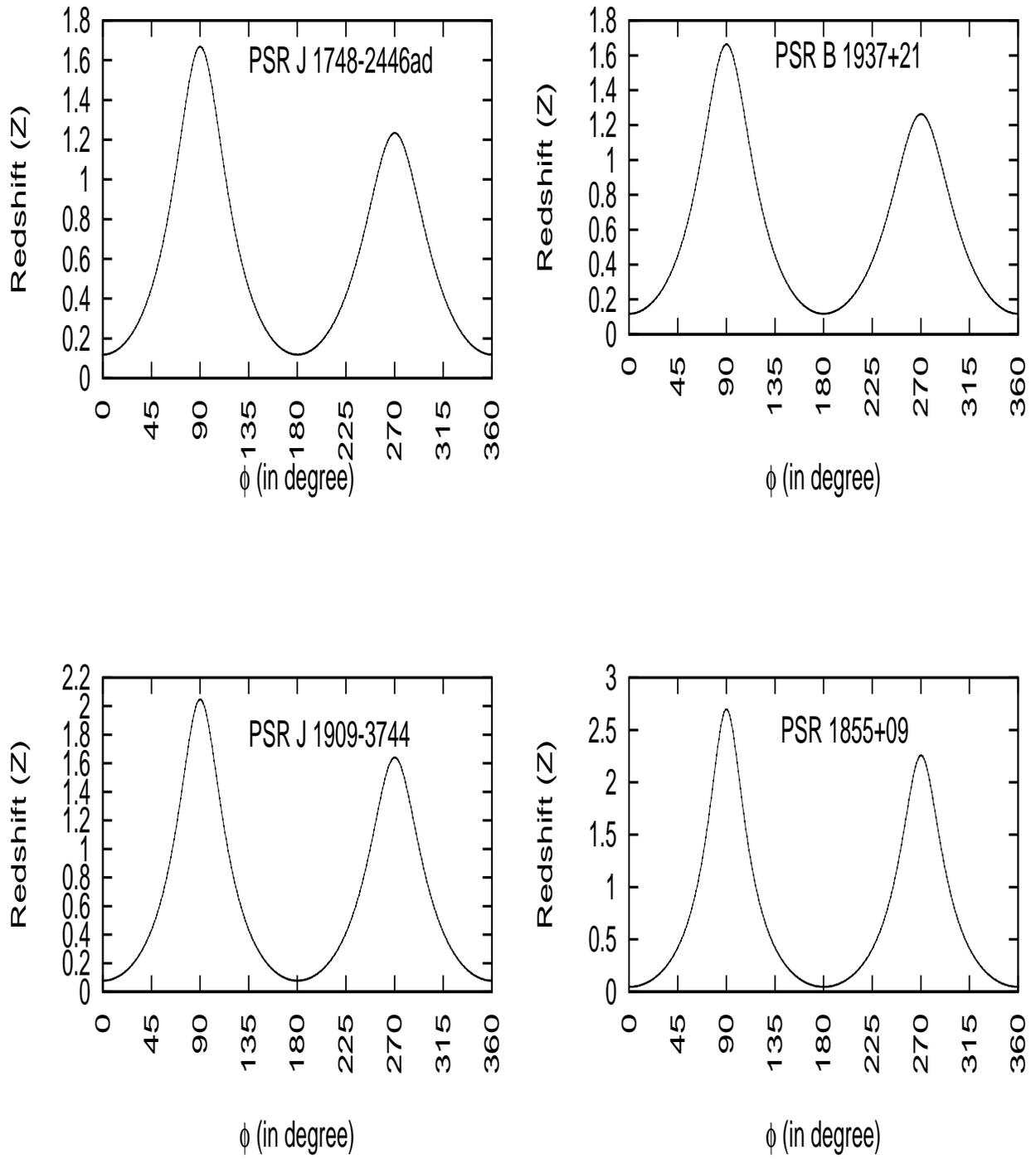}
 \caption{\label{fig1} Shows the variation of redshift (Z) versus $\phi$ for pulsars 1-4 (listed in TABLE - I) from $\phi$ = 0 to $360^o$,  at fixed $\theta$ = $90^o$.}
\end{figure*}
\begin{figure*}
\includegraphics[width=50pc, height=40pc,angle=270]{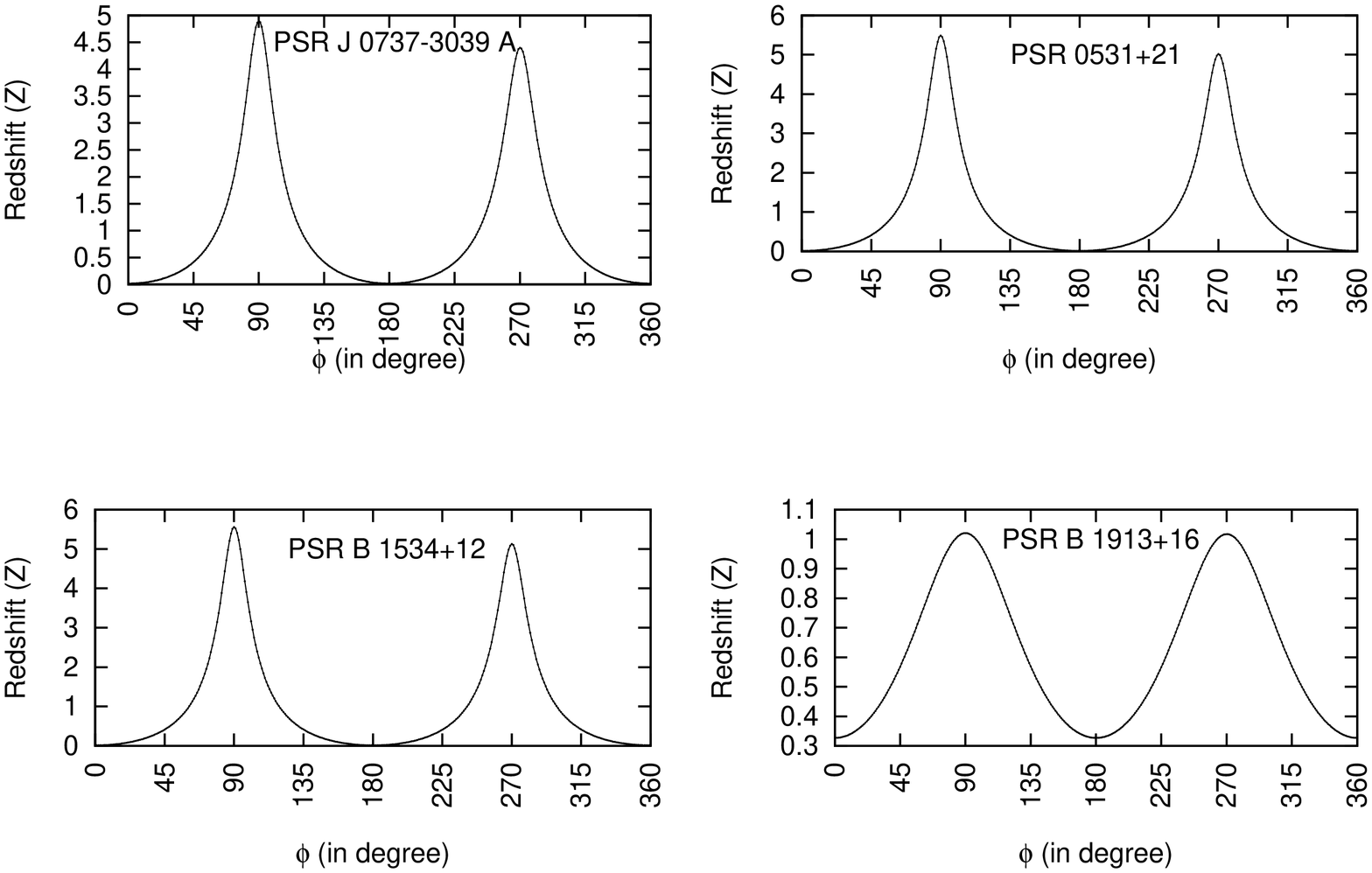}
 \caption{\label{fig2} Shows the variation of redshift (Z) versus $\phi$ for pulsars 5-8 (listed in TABLE - I) from $\phi$ = 0 to $360^o$,  at fixed $\theta$ = $90^o$.}
\end{figure*}
\begin{figure*}
\includegraphics[width=50pc, height=40pc,angle=270]{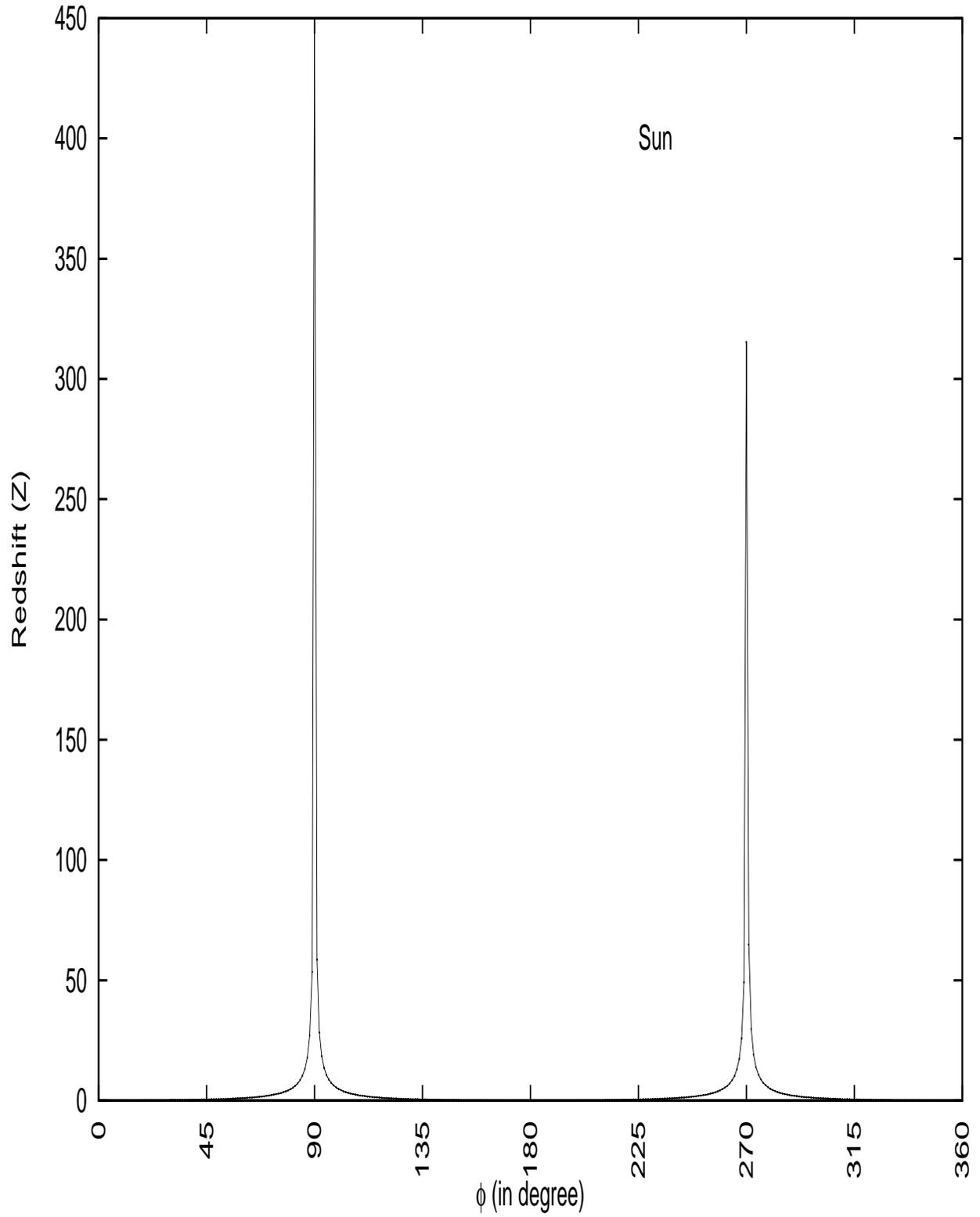}
 \caption{\label{fig3} Shows the variation of redshift (Z) versus $\phi$ for Sun from $\phi$ = 0 to $360^o$,  at fixed $\theta$ = $90^o$.}
\end{figure*}
\begin{table*}
\caption{\label{tab:table1}A list showing  Mass (M), Schwarzschild radius ($r_{g}$), Radius (R), Rotation velocity ($\Omega$), and Rotation parameter (a) for different millisecond pulsars and Sun. Data taken from various authors as cited in column 2.}
\begin{tabular}{|c|c|c|c|c|c|c|}
                      \hline
                      S. No. & Star  & M$(M_{Sun})$ & $r_{g}$(km) & R(km) & $\Omega$(rad/s) & a(km) \\ \hline
                      1 & PSR J 1748-2446ad [17] & 1.350 & 4.050 & 20.1 & 4.4985$\times10^{3}$  & 2.42325   \\
                      2 &PSR B 1937+21 [18]& 1.350 & 4.050 & 20.2 & 4.0334$\times10^{3}$  & 2.19438  \\
                      3 & PSR J 1909-3744 [19] & 1.438 & 4.314 & 31.1 & 2.1300$\times10^{3}$  & 2.74687  \\
                      4 & PSR 1855+09 & 1.350 & 4.050 & 46.9 & 1.1849$\times10^{3}$  & 3.47509   \\
                      5 & PSR J 0737-3039 A [20] & 1.340 & 4.020 & 133.6 & 0.2766$\times10^{3}$ & 6.58269   \\
                      6 & PSR 0531+21 & 1.350 & 4.050 & 164.4 & 0.1885$\times10^{3}$  & 6.79287   \\
                      7 & PSR B 1534+12 [21] & 1.340 & 4.020 & 165.9 & 0.1657$\times10^{3}$  & 6.08070   \\
                      8 & PSR B 1913+16 & 1.440 & 4.320 & 10.0 & 0.1064$\times10^{3}$  & 0.01418  \\
                      9 & Sun & 1.000 & 3.000 & 0.7$\times10^{6}$ & 2.5982$\times10^{-6}$  & 1.69749 \\ \hline
\end{tabular}
\end{table*}
\begin{table*}
\caption{\label{tab:table3}Calculated values of Redshift (Z) for $\phi=0,\frac{\pi}{2} \ and \ \frac{3\pi}{2}$, at fixed $\theta=\frac{\pi}{2}$, corresponding redshifted  wavelength of Lyman - $\alpha$ line (emitted wavelength of 1215.668 ${\AA}$) and the coefficient of latitude dependence of redshift $(\kappa)$ for different millisecond pulsars, in case of (+,+) or (-,-), for (a, $\theta$) combination.}
\begin{tabular}{|c|c|c|c|c|c|c||c||c|}
                      \hline
                      S. No. & Pulsar  & $\beta =\frac{R}{r_{g}}$ & $\gamma=\frac{a}{r_{g}}$ & $Z (\phi =0)$ & Z($\phi =\frac{\pi}{2})$ &$ Z(\phi =\frac{3\pi}{2})$ &$ \ \lambda_{red}(\phi =\frac{\pi}{2})$ & $\kappa (\phi =\frac{\pi}{2},\theta=0^o)$ \\ \hline
                      1 & PSR J 1748-2446ad &4.96296 &0.59833  & 0.118672212 & 1.668503982 &1.233631446 & 3244.0148& 0.418608 \\
                      2 & PSR B 1937+21&4.98765 &0.54182 & 0.118053061 &1.663551032 &1.263311491 &3237.9936 &0.419269\\
                      3 & PSR J 1909-3744 &7.20909 &0.63673 &0.077429231 &2.047082682 &1.640305917 &3704.2408 &0.353403 \\
                      4 & PSR   1855+09 & 11.58025 &0.85805 &0.046167768  &2.697244032  &2.256888342 &4494.6211 &0.282891 \\
                      5 & PSR J 0737-3039 A &33.23383  & 1.63749 &0.015392030  &4.897889501 &4.399200503 &7169.8753 &0.172153 \\
                      6 & PSR   0531+21 &40.59259 &1.67725  &0.012549338 &5.488016592  &5.017013830 &7887.2739 &0.156059 \\
                      7 & PSR B 1534+12 &41.26866  &1.51261 &0.012340052  &5.556341395  & 5.126599154 &7970.3342&0.154401 \\
                      8 & PSR B 1913+16 &2.31481 &0.00328 &0.326861792  &1.020381544  &1.017122955 & 2456.1131&0.656737 \\
                       \hline
\end{tabular}
\end{table*}
In General relativity, redshift (Z) and redshift factor ($\Re$) are defined as [15,16],
\begin{equation}
\frac{1}{Z+1}= \Re =\frac{\omega}{\omega^{'}}
\end{equation}
 A redshift of zero corresponds to an un-shifted line, whereas $Z<0$ indicates blue-shifted emission and $Z>0$ red-shifted emission. A redshift factor of unity corresponds to an un-shifted line, whereas $\Re<1$ indicates red-shifted emission and $\Re>1$ blue-shifted emission.
\\ From  equations (63) and (68), we can write redshift factor ($\Re$) as:

\begin{equation}
\Re(\phi ,\theta)= \sqrt{{g_{tt}+ g_{\phi\phi}(\frac{d\phi}{c dt})^{2}} +2 g_{t \phi}(\frac{d\phi}{c dt})}
\end{equation}

Using equation (68), we can obtain redshift (Z) from above equation (69).
\\If we ignore the electric and magnetic charges contribution, then using the values of metric coefficients $g_{tt}$, $g_{\phi\phi}$, and $g_{t\phi}$ from equations (3), (6) and (7) of Kerr family in equation (69) and substituting the value of $\frac{d\phi}{cdt}$ from equation (46), we can obtain the redshift factor ($\Re$) and corresponding redshift (Z) for the values of $\theta$ from zero (pole) to $\frac{\pi}{2}$ (equator)  and $\phi$ from zero to $2\pi$.\\
\begin{widetext}
$$\Re^{2}(\phi,\theta)=(1-\frac{r_{g} r}{r^{2}+a^{2}cos^{2}\theta})$$
$$ -(r^{2}+a^{2}+\dfrac{r_{g} r a^{2}sin^{2}\theta}{r^{2}+a^{2}cos^{2}\theta})(sin^{2}\theta)(\frac{\frac{R sin\phi sin\theta}{sin^{2}\theta} (1-\frac{r_{g}r}{r^{2}+a^{2}cos^{2}\theta})+\frac{r_{g}ra}{r^{2}+a^{2}cos^{2}\theta}}{-\frac{r_{g}r a}{r^{2}+a^{2}cos^{2}\theta} (R sin \phi sin\theta) + (r^{2}+a^{2}+\frac{r_{g}r a^{2}}{ r^{2}+a^{2}cos^{2}\theta}sin^{2}\theta)})^{2}$$
\begin{equation}
+2 \frac{a sin^{2}\theta (r_{g}r)}{r^{2}+a^{2}cos^{2}\theta}(sin^{2}\theta)(\frac{\frac{R sin\phi sin\theta}{sin^{2}\theta} (1-\frac{r_{g}r}{r^{2}+a^{2}cos^{2}\theta})+\frac{r_{g}ra}{r^{2}+a^{2}cos^{2}\theta}}{-\frac{r_{g}r a}{r^{2}+a^{2}cos^{2}\theta} (R sin \phi sin\theta) + (r^{2}+a^{2}+\frac{r_{g}r a^{2}}{ r^{2}+a^{2}cos^{2}\theta}sin^{2}\theta)})
\end{equation}
\end{widetext}
We can replace \lq r\rq \ by \lq R\rq \ which is the radius of the star.  Thus the redshift clearly depends on the co-ordinates $\theta$ and $\phi$.
 \\ If the observer is on the axis of rotating star, we substitute $\theta =0$ and get the expression for redshift factor as: 
\begin{equation}
\Re(\phi,\theta=0)=\sqrt{(1-\frac{r_{g} r}{r^{2}+a^{2}})}
\end{equation}
When we consider the rotation velocity of central body ($\Omega$) to be zero (or spin parameter, a=0), then we can obtain the corresponding gravitational redshift from a static body of same mass (Schwarzschild Mass).
\\\\ Now at $\theta$ = $\frac{\pi}{2}$ for equatorial plane, substituting the values of $g_{tt}$, $g_{\phi\phi}$, and $g_{t\phi}$ from equations (10), (11) and (12) and $\frac{d\phi}{cdt}(\phi,\theta =\frac{\pi}{2})$ from equation (47) in expression (69) we get:
\begin{widetext}
$$\Re(\phi,\theta =\frac{\pi}{2})$$
\begin{equation}
= \sqrt{(1-\frac{r_{g}}{r}) -(r^{2}+a^{2}+\frac{r_{g}a^{2}}{r})(\frac { (1-\frac{r_{g}}{r})( R sin\phi)+\frac{r_{g}a}{r} }{-\frac{r_{g} a}{r} ( R sin\phi) + (r^{2}+a^{2}+\frac{r_{g} a^{2}}{r})})^{2}+2 (\frac{r_{g}a}{r})(\frac { (1-\frac{r_{g}}{r})( R sin\phi)+\frac{r_{g}a}{r}}{-\frac{r_{g} a}{r} ( R sin\phi) + (r^{2}+a^{2}+\frac{r_{g} a^{2}}{r})})}
\end{equation}
\\Now as the photon is emitted from the surface of the star (Pulsar or Sun), we can substitute the actual value for \lq r\rq, which is \lq R\rq.
$$\Re(\phi,\theta =\frac{\pi}{2})$$
\begin{equation}
= \sqrt{(1-\frac{r_{g}}{R}) -(R^{2}+a^{2}+\frac{r_{g}a^{2}}{R})(\frac { (1-\frac{r_{g}}{R})( R sin\phi)+\frac{r_{g}a}{R} }{-\frac{r_{g} a}{R} ( R sin\phi) + (R^{2}+a^{2}+\frac{r_{g} a^{2}}{R})})^{2}+2 (\frac{r_{g}a}{R})(\frac { (1-\frac{r_{g}}{R})( R sin\phi)+\frac{r_{g}a}{R}}{-\frac{r_{g} a}{R} ( R sin\phi) + (R^{2}+a^{2}+\frac{r_{g} a^{2}}{R})})}
\end{equation}
on simplifying
\begin{equation}
 =\sqrt{\frac{R(R^{4}(R-r_{g})+a^{4}(r_{g}+R)+a^{2}R(2R^{2}- r_{g}^{2})-2aRr_{g}(a^{2}+R(R-r_{g}))sin\phi+R^{2}(r_{g}-R)(a^{2}+R(R-r_{g}))sin^{2}\phi)}{(R^{3}+a^{2}(r_{g}+R)-aRr_{g}sin\phi)^{2}}}
\end{equation}
We also express all the length scales in unit of $r_{g}$, by substituting; $ \frac{R}{r_{g}} = \beta $, and $ \frac{a}{r_{g}}=\gamma$ in above equation (74). Thus after certain simplifications
\begin{equation}
\Re(\phi,\theta =\frac{\pi}{2})=\sqrt{\frac{\beta(-\beta+\beta^{2}+\gamma^{2})((1+\beta)(\beta^{2}+2\gamma^{2})+(\beta-1)\beta^{2}cos2\phi-4\beta\gamma sin\phi)}{2(\beta^{3}+\gamma^{2}+\beta \gamma^{2}-\beta \gamma sin\phi)^{2}}}
\end{equation}
\end{widetext}
In above equation (75) $\beta$ and $\gamma$ are dimensionless quantities and the redshift factor $\Re(\phi,\theta =\frac{\pi}{2})$ will show a periodic nature with respect to $\phi$. \\\\
For carrying out some numerical calculations we consider a Lyman - $\alpha$ line (1215.668 ${\AA}$) emitted from the surface of the star (Pulsar or Sun).
\\For an asymptotic observer, the redshifted ($\lambda_{red}$) Lyman - $\alpha$ line can be calculated as,
\begin{equation}
\lambda_{red} =  \lambda_{em} (Z+1)= \frac{\lambda_{em}}{\Re}
\end{equation}
Further for doing numerical calculations we have considered the Newtonian approximation of angular momentum (J),
\begin{equation}
 J = \frac{2MR^{2}\Omega}{5}
 \end{equation}
So rotation parameter can be written as
 \begin{equation}
 a=\frac{2R^{2}\Omega}{5c}
 \end{equation}
 where $\Omega$ is angular velocity of rotating body  (Pulsar or Sun) as described in Section II A. There are two possibilities for direction of spin ($\pm\Omega$) of central body.
 \\ We have four cases for the rotation parameter (a) and the position ($\theta$) combination on the central body at which photon is emitted;
  \begin{itemize}
    \item a = +ive , $\theta$ = +ive , $\equiv (+,+)$
    \item a = -ive , $\theta$ = -ive , $\equiv (-,-)$
    \item a = +ive , $\theta$ = -ive , $\equiv (+,-)$
    \item a = -ive , $\theta$ = +ive , $\equiv (-,+)$
  \end{itemize}
The Mass (M), Physical radius (R), Schwarzschild radius $r_{g}$,  Rotation velocity ($\Omega$), and Rotation parameter ($a$) of Pulsars and Sun are calculated from different sources in literature and these are listed in TABLE - I.
\\\\ As can be seen from equation (75) each pulsar can be characterized by only two parameters $\beta$ and $\gamma$. In our case, we make TABLE - II, listing $\beta$ and $\gamma$ values of these eight pulsars along  with Z$(\phi =0,\theta =\frac{\pi}{2})$, Z$(\phi =\frac{\pi}{2},\theta =\frac{\pi}{2})$, Z$(\phi =\frac{3\pi}{2},\theta =\frac{\pi}{2})$ and Lyman - $\alpha$ line redshifted wavelength $\ \lambda_{red} (\phi =\frac{\pi}{2},\theta =\frac{\pi}{2})$ values.
\\\\ In FIG.-1 and FIG.-2, we make a plot showing variation of $Z(\phi,\theta =\frac{\pi}{2})$ with $\phi$ for eight pulsars (listed in TABLE - I). We are making plot here for (+,+) or (-,-), for (a, $\theta$) combination. It may be noted that $\phi$ is actually equal to $\Omega t$, where t is the time measured in observers coordinate. The nature is clearly periodic with different amplitudes at $\phi =\frac{\pi}{2}$ and $\frac{3\pi}{2}$. Thus for a pulsar as it rotates the redshifted values of spectrum shows a periodicity. From FIG.-1, FIG.-2 and TABLE - II,  it is clearly seen that the value of redshift (Z) for these millisecond pulsars obtains a maximum (primary maxima) at $ \phi =\frac{\pi}{2},\theta =\frac{\pi}{2}$ and a second maximum (secondary maxima with  value less than the primary maxima) at $ \phi =\frac{3 \pi}{2},\theta =\frac{\pi}{2}$. This asymmetry happens clearly due to the presence of $sin\phi$ and $sin^{2}\phi$  terms in the expression of $\Re$ as in equations (72), (73) and (74). The value of redshift, at $\phi =0,\theta =\frac{\pi}{2}$ is  actually same as that value of gravitational redshift which is generally referred (in most of the published work) as the gravitational redshift in Kerr field. The redshifted wavelength value $ \lambda_{red} (\phi =\frac{\pi}{2},\theta =\frac{\pi}{2})$ is also given in TABLE - II. This calculation have been done for  Lyman - $\alpha$ line. For comparison we make a similar plot (FIG.-3) for Sun and the effect is too small.
\section{\label{sec:Redshifted wavelength} The coefficient of latitude dependence of redshift}
\begin{figure*}
\includegraphics[width=50pc, height=40pc,angle=270]{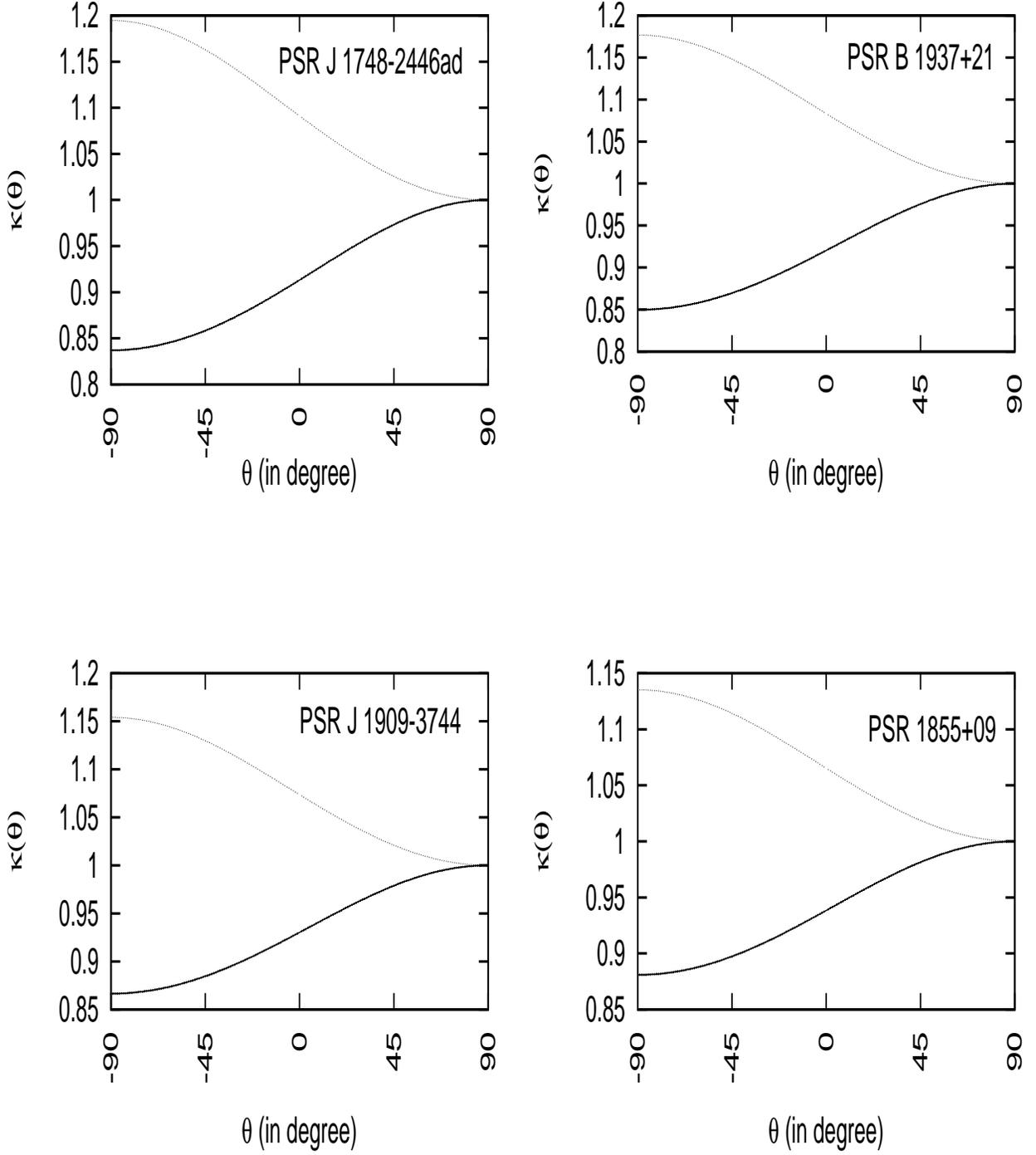}
 \caption{\label{fig4} Shows the variation of the coefficient of latitude dependence of redshift $\kappa (\theta)$  versus $(\theta)$, when  rotation parameter (a) and $\theta$ having same sign (+,+) or (-,-) and opposite sign (+,-) or (-,+) for pulsars 1-4 (listed in TABLE - I) from $\theta =-90^o$ to $90^o$, at fixed $\phi = 90^o$. The solid line represents the case for (+,+) or (-,-) and the dotted line represents the case for (+,-) or (-,+), for (a, $\theta$) combination.}
\end{figure*}
\begin{figure*}
\includegraphics[width=50pc, height=40pc,angle=270]{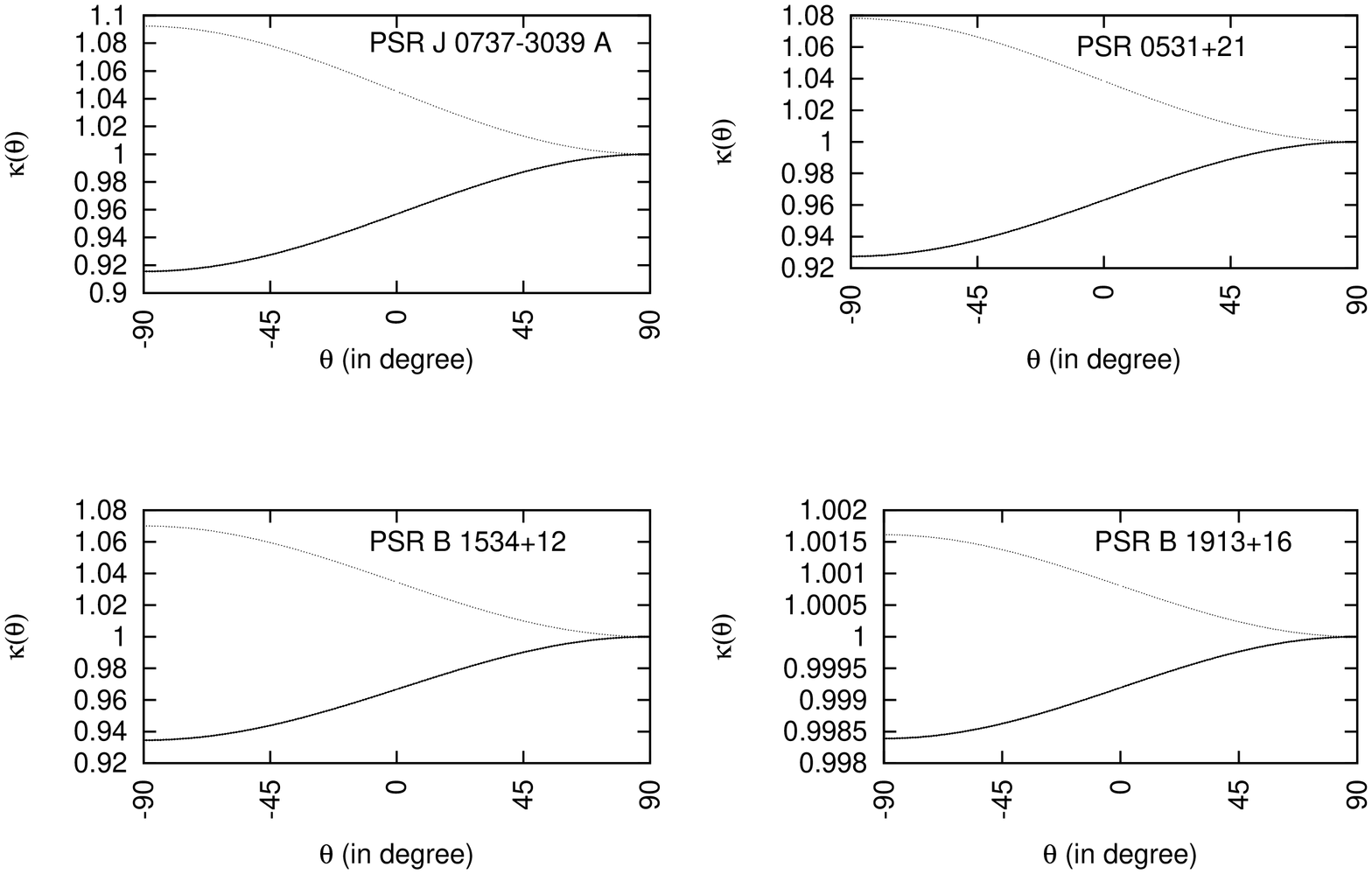}
 \caption{\label{fig5} Shows the variation of the coefficient of latitude dependence of redshift $\kappa (\theta)$  versus $(\theta)$, when  rotation parameter (a) and $\theta$ having same sign (+,+) or (-,-) and opposite sign (+,-) or (-,+) for pulsars 5-8 (listed in TABLE - I) from $\theta =-90^o$ to $90^o$, at fixed $\phi = 90^o$. The solid line represents the case for (+,+) or (-,-) and the dotted line represents the case for (+,-) or (-,+), for (a, $\theta$) combination.}
\end{figure*}
\begin{figure*}
\includegraphics[width=50pc, height=40pc,angle=270]{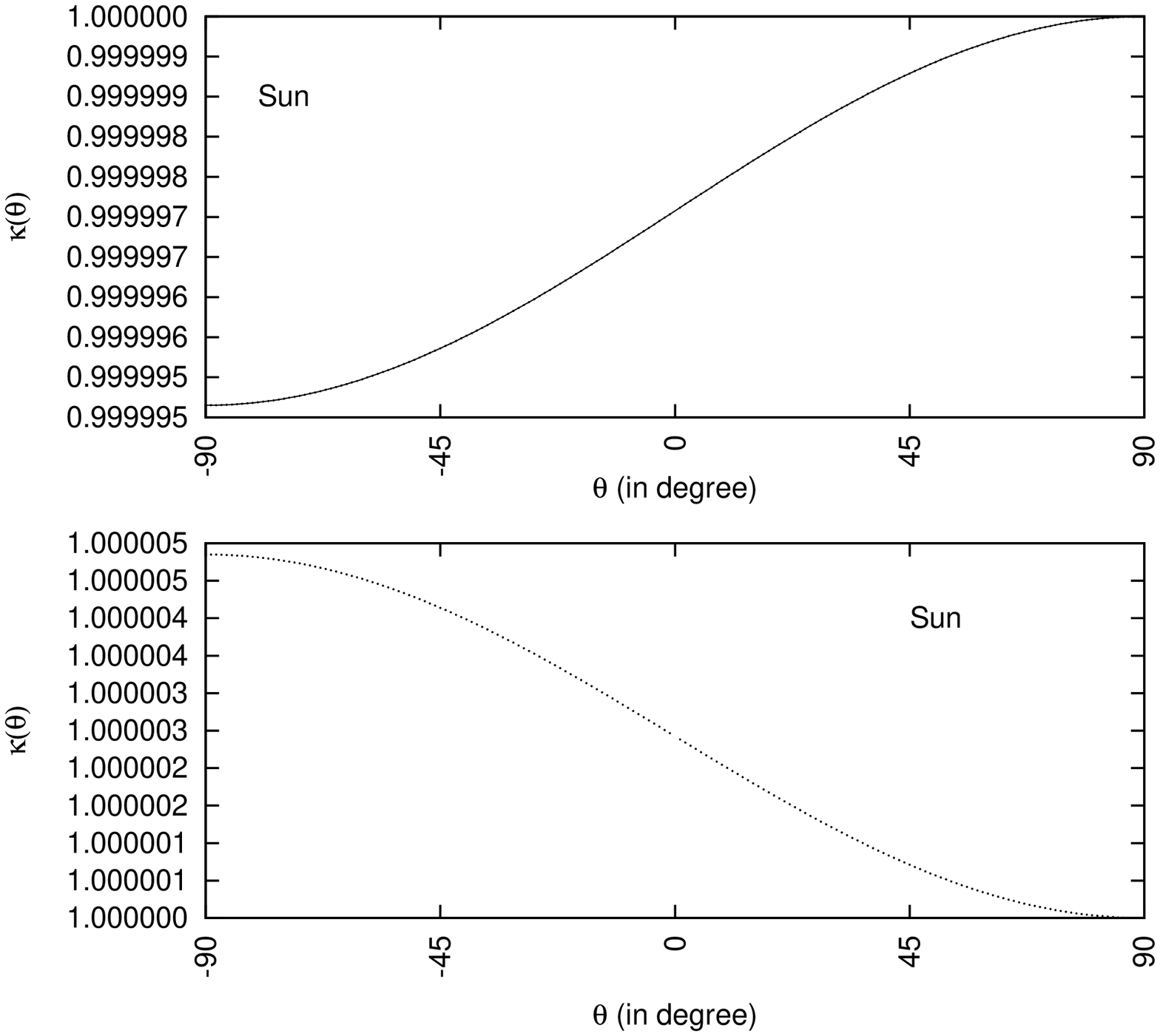}
 \caption{\label{fig6} Shows the variation of the coefficient of latitude dependence of redshift $\kappa (\theta)$  versus $(\theta)$, when  rotation parameter (a) and $\theta$ having same sign (+,+) or (-,-) and opposite sign (+,-) or (-,+) for Sun from $\theta =-90^o$ to $90^o$, at fixed $\phi = 90^o$. The solid line represents the case for (+,+) or (-,-) and the dotted line represents the case for (+,-) or (-,+), for (a, $\theta$) combination.}
\end{figure*}
Below we define a parameter $\kappa(\phi=\frac{\pi}{2},\theta)$, showing latitude dependence of redshift.
\\ From equation (76), we can write
\begin{widetext}
\begin{equation}
\frac{\lambda_{red, \phi=\frac{\pi}{2},\theta}}{\lambda_{red, \phi=\frac{\pi}{2},\theta=\frac{\pi}{2}}} = \frac{(Z+1)_{ \phi=\frac{\pi}{2},\theta}}{(Z+1)_{ \phi=\frac{\pi}{2},\theta=\frac{\pi}{2}}} =\frac{\Re_{ \phi=\frac{\pi}{2},\theta =\frac{\pi}{2}}}{\Re_{ \phi=\frac{\pi}{2},\theta}}=\kappa(\phi=\frac{\pi}{2},\theta) (say)
\end{equation}
The quantity $\kappa(\phi=\frac{\pi}{2},\theta)$ thus, is defined as red-shifted wavelength ratio at any value of latitude on the central body (star) at which photon is emitted and at equator. This parameter can be termed as \lq coefficient of latitude  dependence of redshift from a rotating star\rq. \\
Using the value of $\Re$ from equation (69) in above equation (79), we can write
\begin{equation}
 \kappa(\phi=\frac{\pi}{2},\theta) = \frac{[\sqrt{{g_{tt}+ g_{\phi\phi}(\frac{d\phi}{c dt})^{2}} +2 g_{t \phi}(\frac{d\phi}{c dt})}] _ {\theta=\frac{\pi}{2}}}{[\sqrt{{g_{tt}+ g_{\phi\phi}(\frac{d\phi}{c dt})^{2}} +2 g_{t \phi}(\frac{d\phi}{c dt})}]_ {\theta}}
 \end{equation}
Here we can substitute the value of $\frac{d\phi}{cdt}(\phi =\frac{\pi}{2},\theta )$  from equation (48) and $g_{tt}$, $g_{\phi\phi}$ and $g_{t\phi}$ from equations (3), (6) and (7) for any value of $\theta$.  We can also substitute the value of $\frac{d\phi}{cdt}(\phi =\frac{\pi}{2},\theta =\frac{\pi}{2})$ from equation (49) and $g_{tt}$, $g_{\phi\phi}$ and $g_{t\phi}$ from equations (10), (11) and (12) for $\theta =\frac{\pi}{2}$.\\
As a result we can write $$\kappa(\phi=\frac{\pi}{2},\theta)=$$
 \begin{equation}
 \frac{\sqrt{{(1-\frac{r_{g}}{r}) -(r^{2}+a^{2}+\frac{r_{g}a^{2}}{r})(\frac{R (1-\frac{r_{g}}{r})+\frac{r_{g}a}{r}}{-\frac{r_{g} a}{r} R + (r^{2}+a^{2}+\frac{r_{g} a^{2}}{ r})})^{2}} + 2(\frac{r_{g}a}{r})(\frac{R (1-\frac{r_{g}}{r})+\frac{r_{g}a}{r}}{-\frac{r_{g} a}{r} R + (r^{2}+a^{2}+\frac{r_{g} a^{2}}{ r})})}}{\sqrt{{(1-\frac{r_{g} r}{\rho^{2}})-[r^{2}+a^{2}+\dfrac{r_{g} r a^{2}sin^{2}\theta}{\rho^{2}}]sin^{2}\theta(\frac{\frac{R}{sin\theta} (1-\frac{r_{g}r}{\rho^{2}})+\frac{r_{g}ra}{\rho^{2}}}{-\frac{r_{g}r a}{\rho^{2}} (R sin\theta) + (r^{2}+a^{2}+\frac{r_{g}r a^{2}}{ \rho^{2}}sin^{2}\theta)})^{2}} +2 (\frac{a sin^{2}\theta r_{g}r}{\rho^{2}})(\frac{\frac{R}{sin\theta} (1-\frac{r_{g}r}{\rho^{2}})+\frac{r_{g}ra}{\rho^{2}}}{-\frac{r_{g}r a}{\rho^{2}} (R sin\theta) + (r^{2}+a^{2}+\frac{r_{g}r a^{2}}{ \rho^{2}}sin^{2}\theta)})}}
 \end{equation}
 \end{widetext}
 The denominator of the fraction in above equation (81) is redshift factor $\Re(\phi=\frac{\pi}{2},\theta)$, which clearly depends on the direction of spin ($\pm a$) of central body and also on the position ($\pm\theta$) on the central body at which the photon is emitted.
 \\ This also indicates that $\kappa(\phi=\frac{\pi}{2},\theta)$ will also depend on the position $(\pm\theta)$ and rotation parameter ($\pm$ a).
  \\\\ We obtain same results for redshift (Z) and $\kappa(\phi=\frac{\pi}{2},\theta)$ in case of (+,+) or (-,-) and similarly the same results in case of (+,-) or (-,+), for (a, $\theta$) combination.
 \\ Finally we have only two situations for the combination of rotation parameter (a) and the position ($\theta$) on the central body at which photon is emitted; (+,+) or (-,-) and (+,-) or (-,+), for (a, $\theta$) combination.
\\\\ We can find $\kappa(\phi=\frac{\pi}{2},\theta)$ for $\theta =0$, which will give a ratio of redshifted wavelength between pole and equatorial region.
\begin{widetext}
\begin{equation}
\kappa(\phi=\frac{\pi}{2},\theta=0) =\frac{[\sqrt{{g_{tt}+ g_{\phi\phi}(\frac{d\phi}{c dt})^{2}} +2 g_{t \phi}(\frac{d\phi}{c dt})}] _ {\theta=\frac{\pi}{2}}}{[\sqrt{{g_{tt}+ g_{\phi\phi}(\frac{d\phi}{c dt})^{2}} +2 g_{t \phi}(\frac{d\phi}{c dt})}]_ {\theta=0}}
\end{equation}
 At poles ($\theta=0$), we use the values $g_{tt} =(1-\frac{r_{g}r}{r^{2}+a^{2}})$ and  $g_{\phi\phi}=g_{t\phi}=0$, from equations (13) to (15). Therefore, equation (82) can be rewritten as
\begin{equation}
\kappa(\phi=\frac{\pi}{2},\theta=0) =\frac{[\sqrt{{g_{tt}+ g_{\phi\phi}(\frac{d\phi}{c dt})^{2}} +2 g_{t \phi}(\frac{d\phi}{c dt})}]_ {\theta=\frac{\pi}{2}}}{\sqrt{(1-\frac{r_{g}r}{r^{2}+a^{2}}})}
\end{equation}

Substituting the value of $\frac{d\phi}{cdt}(\phi =\frac{\pi}{2},\theta =\frac{\pi}{2})$ from equation (49) and $g_{tt}$, $g_{\phi\phi}$ and $g_{t\phi}$ from equations (10), (11) and (12) for $\theta=\frac{\pi}{2}$, above equation (83) can be written as

\begin{equation}
\kappa(\phi=\frac{\pi}{2},\theta=0) =\frac{\sqrt{(1-\frac{r_{g}}{r})-(r^{2}+a^{2}+\frac{r_{g}a^{2}}{r})(\frac{R (1-\frac{r_{g}}{r})+\frac{r_{g}a}{r}}{-\frac{r_{g} a}{r} R + (r^{2}+a^{2}+\frac{r_{g} a^{2}}{ r})})^{2}+ 2(\frac{r_{g}a}{r})(\frac{R (1-\frac{r_{g}}{r})+\frac{r_{g}a}{r}}{-\frac{r_{g} a}{r} R + (r^{2}+a^{2}+\frac{r_{g} a^{2}}{ r})})}}{\sqrt{(1-\frac{r_{g}r}{r^{2}+a^{2}}})}
\end{equation}
\end{widetext}
In FIG.-4 and 5, we show $\kappa(\phi=\frac{\pi}{2},\theta)$ for eight pulsars listed in TABLE - I and II as a function of $\theta$. It is  seen that the value of $\kappa(\phi=\frac{\pi}{2},\theta)$  is less than one for (+,+) or (-,-) combination while greater than one for (+,-) or (-,+), for (a, $\theta$) combination. This clearly indicates that the gravitational redshift  increases from pole to equatorial region (maximum at equator). In case of (+,+) or (-,-), for (a, $\theta$) combination. Again it decreases from pole to equatorial region (minimum at equator), in case of (+,-) or (-,+), for (a, $\theta$) combination. This shows clearly the dependence of redshift on the latitude ($90^o - \theta$). The values of $\kappa(\phi =\frac{\pi}{2},\theta=0^o)$ for these pulsars are also given in TABLE - II, for (+,+) or (-,-) combination. We make FIG.-6, for Sun showing $ \kappa(\theta)$ versus $\theta$, for the purpose of comparison and we notice that the effect is too small.

\section{\label{sec:Concl}Conclusions}
 We can conclude from the present work that,
\begin{enumerate}
  \item  Gravitational redshift is affected by rotation of the central body.
  \item Gravitational redshift is a function of the $\phi$ and latitude ($90^o - \theta$).
  \item Gravitational redshift will depend on the direction of spin ($\pm a$) of central body and also on the position ($\pm\theta$) on the central body at which the photon is emitted.
  \item When we consider the rotation velocity of central body ($\Omega$) to be zero, then we can obtain the corresponding gravitational redshift from a static body of same mass (Schwarzschild Mass).
  \item Gravitational redshift shows a periodic nature with respect to $\phi$. The value of redshift (Z) obtains a maximum (primary maxima) at, $ \phi =\frac{\pi}{2},\theta =\frac{\pi}{2}$ and a second maximum (secondary maxima with  value less than the primary maxima) at $ \phi =\frac{3 \pi}{2},\theta =\frac{\pi}{2}$.
  \item Gravitational redshift  increases from pole to equatorial region (maximum at equator), when  rotation parameter (a) and the position ($\theta$) on the central body at which the photon is emitted having same signs (+,+) or (-,-), for (a, $\theta$) combination.
  \item  Gravitational redshift decreases from pole to equatorial region (minimum at equator), when  rotation parameter (a) and the position ($\theta$) on the central body at which the photon is emitted having opposite signs (+,-) or (-,+), for (a, $\theta$) combination.
  \item The coefficient of latitude dependence of redshift ($\kappa$), when the  photon is emitted from any value of latitude on the central body and equator, clearly shows the dependence of redshift on the latitude ($90^o - \theta$).
  \item Looking at the observed redshift from a rotating object it should be possible to comment, from what latitude ($90^o - \theta$) of the object, the Lyman - $\alpha$  line has been radiated.
\end{enumerate}
\begin{acknowledgments}
 We wish to thank Dr. Atri Deshmukhya, Head  Department of Physics, Assam University, Silchar, India for inspiring discussions. A K Dubey is also thankful to Sanjib Deb and Arindwam Chakraborty, Department of Physics, Assam University, for providing help and support in programming and plots.
\end{acknowledgments}

\end{document}